\documentclass[epj]{webofc}
\usepackage[utf8]{inputenc}
\usepackage[varg]{txfonts}   
\usepackage{booktabs}
\usepackage{xcolor}
\definecolor{darkred}{rgb}{0.4,0.0,0.0}
\definecolor{darkgreen}{rgb}{0.0,0.4,0.0}
\definecolor{darkblue}{rgb}{0.0,0.0,0.4}
\usepackage[bookmarks,linktocpage,colorlinks,
    linkcolor = darkred,
    urlcolor  = darkblue,
    citecolor = darkgreen]{hyperref}
\usepackage{subfigure}

\usepackage{graphicx}
\usepackage{amssymb}
\usepackage{amsmath}
\usepackage{epstopdf}
\usepackage{bm}
\usepackage{color}
\usepackage{bbold}
\usepackage{capt-of}
\usepackage{pstricks}
\usepackage{bm} 
\usepackage{slashed}
\usepackage{tikz}

\long\def\comment#1{ }

\newcommand{\beq}{\begin{eqnarray}}
\newcommand{\eeq}{\end{eqnarray}}

\newcommand{\bea}{\begin{eqnarray}}
\newcommand{\eea}{\end{eqnarray}}
\newcommand{\bean}{\begin{eqnarray*}}
\newcommand{\eean}{\end{eqnarray*}}

\newcommand{\pv}{{\mathbf p}}

\newcommand{\half}{\frac{1}{2}}

\newcommand{\tr}{\mbox{tr}\,}
\newcommand{\diag}{\mbox{diag}\,}

\newcommand{\avg}[1]{\left\langle #1 \right\rangle}

\newcommand{\be}{\begin{eqnarray}}
\newcommand{\ee}{\end{eqnarray}}

\renewcommand{\P}{\mathcal{P}}

\newcommand{\id}{\mathbb{1}}

\wocname{EPJ Web of Conferences}
\woctitle{Lattice2017}
\begin{document}
%
\selectlanguage{english}
\title{%
Medium effects and parity doubling of hyperons across the deconfinement phase transition\thanks{Presented at $35^{\rm th}$ International Symposium on Lattice Field Theory, $18$-$24$ June $2017$, Granada, Spain}
}
\author{%
\firstname{Gert} \lastname{Aarts}\inst{1}\and
\firstname{Chris} \lastname{Allton}\inst{1} \and
\firstname{Davide}  \lastname{De Boni}\inst{1}\fnsep\thanks{Speaker, \email{d.de-boni.840671@swansea.ac.uk}} \and
\firstname{Simon} \lastname{Hands}\inst{1} \and
\firstname{Benjamin} \lastname{J\"ager}\inst{2,3} \and
\firstname{Chrisanthi} \lastname{Praki}\inst{1} \and
\firstname{Jon-Ivar} \lastname{Skullerud}\inst{4,5}
}
\institute{%
Department of Physics, College of Science, Swansea University, Swansea SA2 8PP, United
Kingdom
\and
ETH Zürich, Institute for Theoretical Physics, Wolfgang-Pauli-Str. 27, 8093 Zürich, Switzerland
\and
$\mbox{CP}^3$-Origins \& Danish Institute for advanced study, Department of Mathematics and Computer Science, University of Southern Denmark, 5230 Odense M, Denmark
\and
Department of Theoretical Physics, Maynooth University,
County Kildare, Ireland
\and
School of Mathematics, Trinity College Dublin, Dublin 2, Ireland
}
\abstract{%
We analyse the behaviour of hyperons with strangeness $S=-1$,$-2$,$-3$ in the hadronic and quark gluon plasma phases, with particular interest in parity doubling and its emergence as the temperature grows. This study uses our FASTSUM anisotropic $N_f\!\!=$~2+1 ensembles, with four temperatures below and four above the deconfinement transition temperature, $T_c$. The positive-parity groundstate masses are found to be largely temperature independent below $T_c$, whereas the negative-parity ones decrease considerably as the temperature increases. Close to the transition, the masses are almost degenerate, in line with the expectation from chiral symmetry restoration. This may be of interest for heavy-ion phenomenology. In particular we show an application of this effect to the Hadron Resonance Gas model. A clear signal of parity doubling is found above $T_c$ in all hyperon channels, with the strength of the effect depending on the number of $s$-quarks in the baryons.
}
\maketitle
\section{Introduction}\label{intro}

In nature, at zero temperature there is a large difference in mass ($\sim 500$ MeV $>\Lambda_{_{\rm QCD}}$) between the negative- and positive-parity baryonic groundstates.
It is well known that this large mass gap is due to the spontaneous breaking of chiral symmetry, and not only to the small explicit breaking of chiral symmetry coming from the masses of the light quarks.
Chiral symmetry is expected to be restored above the deconfinement temperature, in the quark-gluon plasma phase. The restoration
of chiral symmetry would then imply (in the limit of massless quarks) a degeneracy in the negative- and positive-parity baryonic channels, that is \textit{parity doubling}.
While in lattice QCD there are studies of chiral symmetry at finite temperature in the mesonic sector (an overview can be found in \cite{Aarts:2017rrl}), only a few quenched analyses
are available in the baryonic sector ~\cite{DeTar:1987xb,DeTar:1987ar,Pushkina:2004wa,Datta:2012fz}. 
Here we study parity doubling in the baryonic sector using dynamical lattice QCD simulations, in particular for strange baryons (hyperons).
We analyse the baryonic correlators in both the octet and decuplet sectors, below and above the crossover temperature $T_c$.
Correlators and spectral functions (comprising also the nucleon and $\Delta$-baryon) can be found in the previous works \cite{Aarts:2015mma,Aarts:2015xua,Aarts:2016kvt,Aarts:2017rrl}.

\section{Theoretical aspects of parity doubling}\label{sec-1}

Here we give a short presentation on how parity doubling manifests itself at the level of the baryonic correlators \cite{Aarts:2017rrl}. We exploit charge-conjugation symmetry of the correlator at vanishing chemical potential.
We consider two-point functions of baryonic operators, of the form 
\be
G^{\alpha\alpha'}(x) = \avg{O^\alpha(x)\,\overline{O}^{\alpha'}(0)},
\ee
where $\alpha, \alpha'$ are Dirac indices and $\overline O = O^\dagger \gamma_4$ is a generic baryonic creation operator.
Under parity, the baryonic operators transform as elementary quark fields, that is
$\P O(x)\P^{-1} = \gamma_4O(Px)\,$, with $P=\diag(-1,-1,-1,1)\,$.
Hence one may introduce parity projectors and operators via
\be
\label{eq:Ppm}
 P_{\pm} = \half \left(\id \pm \gamma_4\right), 
 \qquad\quad
 O_\pm(x) = P_\pm O(x),\qquad\Rightarrow\qquad \P O_\pm(x)\P^{-1} = \pm O_\pm(Px).
\ee
We refer to $O_\pm$ as positive- and negative-parity operators. The corresponding positive- and negative-parity correlators are
\be
G_\pm(x) = \tr P_\pm G(x) = \tr \avg{O_\pm(x)\,\overline{O}_\pm(0)}\,. 
\ee
Under charge conjugation the baryonic operators transform again as quark fields, i.e.\
$\!O^{(c)} = C^{-1}\overline O^{^T},\:
 \overline O^{(c)} = -O^{^T}C\,$,
where $C$ corresponds to the charge conjugation matrix.
At zero chemical potential, thermal expectation values are invariant under charge conjugation, from which it follows that \cite{Aarts:2017rrl} 
\be
\label{eq:Gpmmp}
G_\pm(\tau,\pv) = -G_\mp(1/T-\tau,\pv),
\ee
where we used the cyclic property of the trace and isotropy.
This means that positive-parity states propagate forward in euclidean time, when using $G_+$, and backward in time when using  $G_-$, and vice versa for negative-parity states. If the groundstates masses $m_\pm$ are the dominant contribution to the correlator, one has
\be
\label{eq:Gsimple}
\pm G_\pm(\tau)= A_{\pm} e^{-m_\pm\tau} + A_\mp e^{-m_\mp(1/T-\tau)},
\ee
assuming vanishing momentum $\pv=0$. One can show \cite{Gattringer:2010zz,Aarts:2017rrl} that when chiral symmetry is unbroken, i.e.\ the quarks are massless and chiral symmetry is not broken spontaneously, the theory is unchanged under chiral rotations on the quark fields. This implies that
\be
G_+(\tau,\pv) = -G_-(\tau,\pv) \stackrel{(\ref{eq:Gpmmp})}{=} G_+(1/T-\tau,\pv),
\ee
hence the positive- and negative-parity correlators are symmetric with respect to reversal about $\tau=1/(2T)\,$, meaning that the parity groundstate partners have the same mass in the confined phase.

\section{Lattice setup}\label{sec:setup}

\begin{table}[thb]
\small
\centering
\caption{Ensembles used in this work. The lattice size is $N_s^3 \times N_\tau$ , with the temperature $T = 1/(a_\tau N_\tau)$.
The available statistics for each ensemble is $N_{\rm cfg} \times N_{\rm src}$. The sources were placed randomly in the $4$D lattice.
}
\label{tab:lat}
\begin{tabular}{cccccccc}
\hline
$N_s$ & $N_\tau$  & $T$\,[MeV] & $T/T_c$  & $N_{\rm src}$  & $N_{\rm cfg}$\\
\hline
  24 & 128 & 44	& 0.24  &  16 & 139\\
  24 & 40 & 141 & 0.76  &  4  & 501  \\
  24 & 36 & 156 & 0.84  &  4  & 501  \\
  24 & 32 & 176 & 0.95  &  2  & 1000 \\ 
  24 & 28 & 201 & 1.09  &  2  & 1001 \\ 
  24 & 24 & 235 & 1.27  &  2  & 1001 \\  
  24 & 20 & 281 & 1.52  &  2  & 1000 \\ 
  24 & 16 & 352 & 1.90  &  2  & 1001 \\ 
\hline
\end{tabular}
\end{table}

Baryonic correlators are computed using the thermal ensembles of the FASTSUM collaboration \cite{Amato:2013naa,Aarts:2014nba,Aarts:2014cda}. They are generated with $2+1$ flavours of Wilson fermion on an anisotropic lattice, with $a_\tau<a_s$. The spatial lattice spacing is $a_s = 0.1227(8)$ fm and the renormalised anisotropy $\xi\equiv a_s/a_\tau =3.5$.
We used the Symanzik-improved anisotropic gauge action with tree-level mean-field coefficients and a mean-field-improved Wilson-clover fermion action with stout-smeared links, following the Hadron Spectrum Collaboration \cite{Edwards:2008ja}. More details of the action and parameter values can be found in Refs.\ \cite{Aarts:2014cda,Aarts:2014nba}.
The parameters for the degenerate $u$ and $d$ quarks result in a pion mass of $M_\pi=384(4)$ MeV \cite{Lin:2008pr}, a value larger than the one found in nature. The strange quark has been tuned to its physical value.
The CHROMA software package \cite{Edwards:2004sx} allowed us to generate configurations and correlation functions, using the SSE
optimizations when possible \cite{McClendon}.

We use a fixed-scale approach, in which the temperature is varied by changing $N_\tau$, according to $T = 1/(a_\tau N_\tau)$. 
Table \ref{tab:lat} provides a list of the ensembles (four ensembles in the hadronic phase and four in the quark-gluon plasma).
The pseudo-critical temperature has been estimated from an analysis of the renormalized Polyakov loop  \cite{Aarts:2014nba} and
is higher than in nature, due to the large pion mass.

Gaussian smearing \cite{Gusken:1989ad} has been employed on the baryonic correlators to increase the overlap
with respect to the groundstate at the lowest temperature. In order to have a positive spectral weight, we apply the
smearing on both source and sink, i.e., 
$\psi' = \frac{1}{A} \left(\id + \kappa H \right)^{n} \psi\,$,
where $A$ is a normalisation factor and $H$ is the spatial hopping part of the Dirac operator.
The hopping term contains APE smeared links \cite{Albanese:1987ds} using $\alpha =1.33$ with one iteration. 
We use $n=60$ and $\kappa = 4.2$, which maximise
the length of the plateau for the effective mass of the groundstate at the lowest temperature. 
Smearing is applied only in the spatial directions, equally to all temperatures and ensembles.


\section{In-medium effects for hyperons}\label{sec:correlators}

\begin{figure}[t!] 
  \centering
  \includegraphics[width=7cm,clip]{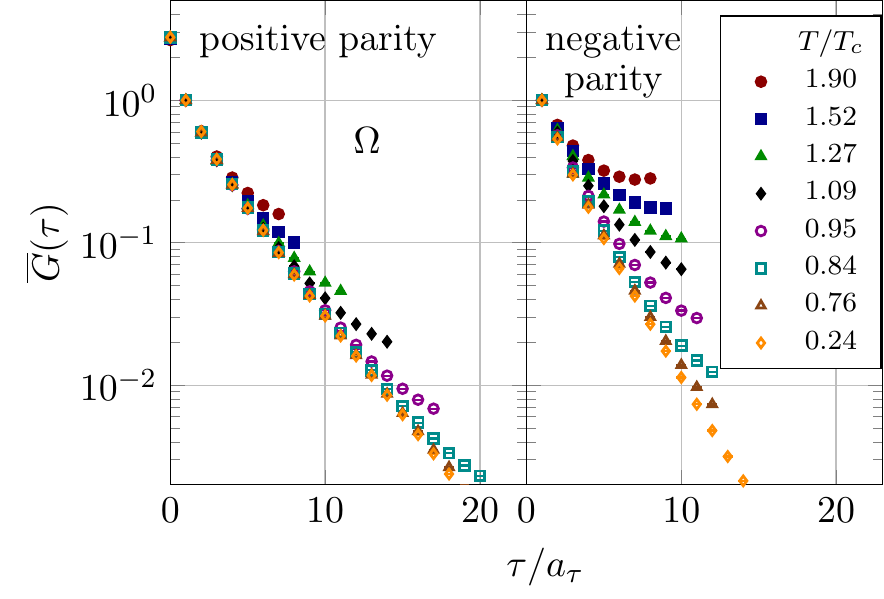}
  \includegraphics[width=6.45cm,clip]{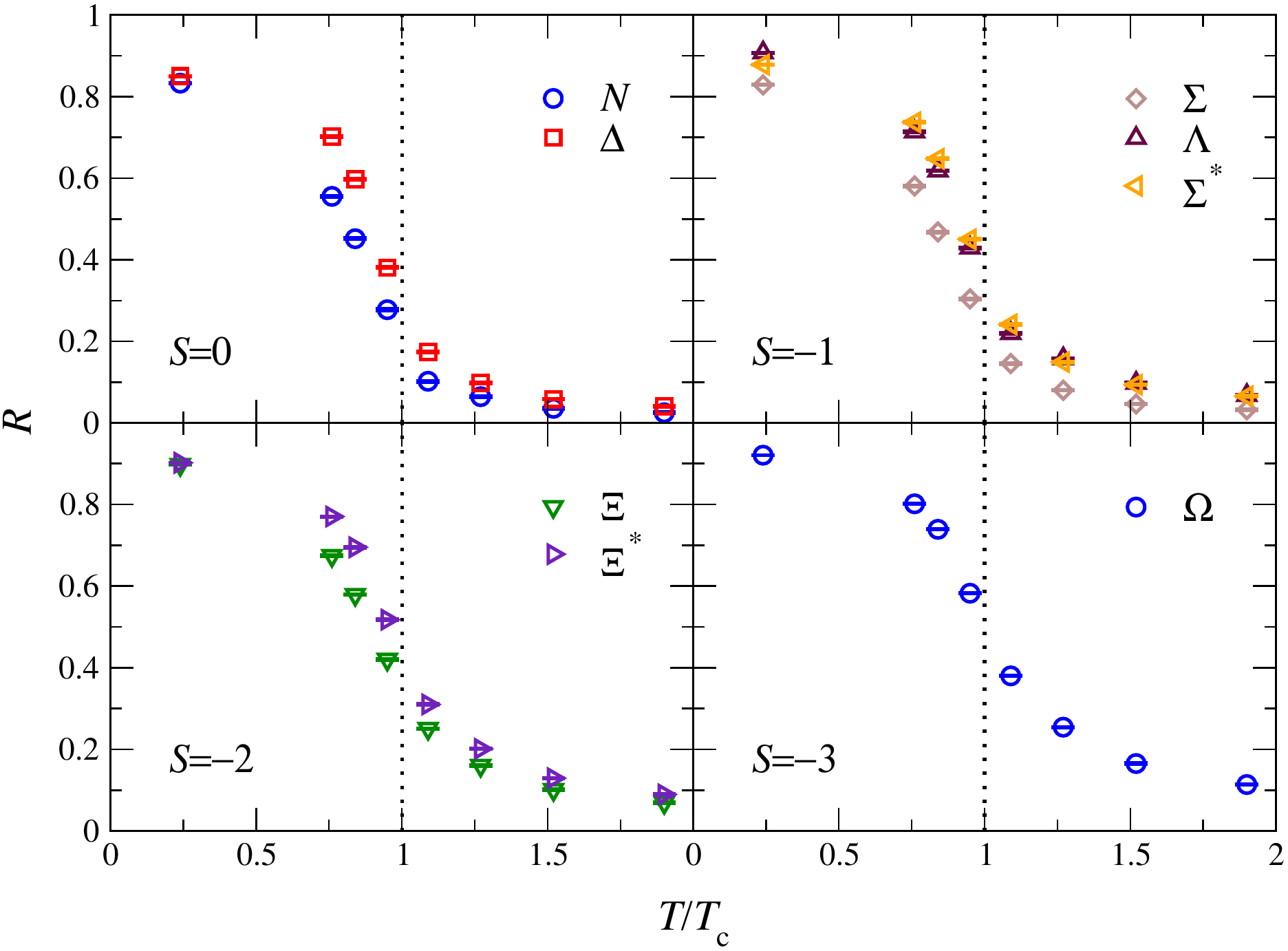}
  \caption{Left: Correlator of the $\Omega$ baryon ($S=-3$) for the positive- and negative-parity channels at different temperatures below and above $T_c\,$. Right: Temperature dependence of the $R$ ratio of all baryons in the spin $1/2$ octet and spin $3/2$ decuplet.}
  \label{fig-1}
\end{figure}

The correlation functions of all hyperons ($\Sigma$, $\Sigma^*$, $\Lambda$, $\Xi$, $\Xi^*$, $\Omega$) of the spin-$1/2$ baryon octet and spin-$3/2$ baryon decuplet have been computed for both parity partners. The left panel of Fig.\ref{fig-1} displays the correlation function of the $\Omega$ baryon, the correlators for the other strange baryons will appear in a subsequent work in  preparation. Their qualitative behaviour is similar to the one of the $\Omega$ baryon. 
In order to better compare all data at different temperatures we
normalized the correlation function to the first Euclidean time $\tau=a_\tau$
($\tau=N_\tau a_\tau -a_\tau$) for the positive-(negative-)parity partner, that is (note that $G_+$ contains information on both parity channels according to eq.(\ref{eq:Gpmmp}))
\be
 \overline{G}_{+}(\tau) = \frac{G_+\left( \tau \right)}{\langle G_+(a_\tau) \rangle} \quad
  \mbox{and} \quad
  \overline{G}_{-}(\tau) = \frac{G_+\left( N_\tau a_\tau- \tau \right)}{\langle G_+(N_\tau a_\tau - a_\tau)
  \rangle}.
\ee
It is clear from Fig.\ref{fig-1} that the negative-parity channel is much more sensitive to the temperature than the positive-parity one. The negative-parity correlator exhibits a steeper decay than the positive-parity one at low temperature, indicating a heavier mass. However, as temperature is increased, the slope of the negative-parity correlator reduces, signalling that the mass of the negative-parity groundstate is decreasing. On the other hand, the slope of the positive-parity channel is much less affected by temperature, suggesting that the mass of the positive-parity groundstate has very little dependence on the temperature of the medium.
 
The right panel of Fig.\ref{fig-1} shows the summed $R$ ratios of all the hyperons plus the nucleon and $\Delta$ baryon, which are defined as
\be
\label{eq:ratioR}
R \equiv
\frac{\sum_{n=1}^{N_\tau/
2-1}R(\tau_n)/\sigma^2(\tau_n)}{\sum_{n=1}^{N_\tau/2-1}1/\sigma^2(\tau_n)}\:,\:\:\mbox{with}\quad R(\tau)\equiv \frac{G_+(\tau)-G_+(1/T-\tau)}{G_+(\tau)+G_+(1/T-\tau)}\:.
\ee
By construction the ratio $R$ lies between $0$ and $1$. It vanishes when the correlator is symmetric with respect to $1/(2T)\,$, that is in the case of parity doubling. In absence of parity doubling and with clearly separated
groundstates, i.e.\ $m_-\gg m_+$, one has instead $R=1\,$. The statistical uncertainties are used as weights in the ratio $R$. The plot shows that around the crossover temperature the parameter $R$ has a significant drop, which is smaller for large strangeness $|S|$. 
Note that the value of $R$ at our largest temperature increases with strangeness. This can be understood since the strange quark mass breaks chiral symmmetry explicitly, but its
effect is expected to disappear eventually as $m_s/T\to 0\,$. This means that the emergence of
parity doubling due to chiral symmetry restoration in the baryonic sector is indeed valid.

\begin{figure}[t!]
  \centering
  \includegraphics[width=6.6cm,clip]{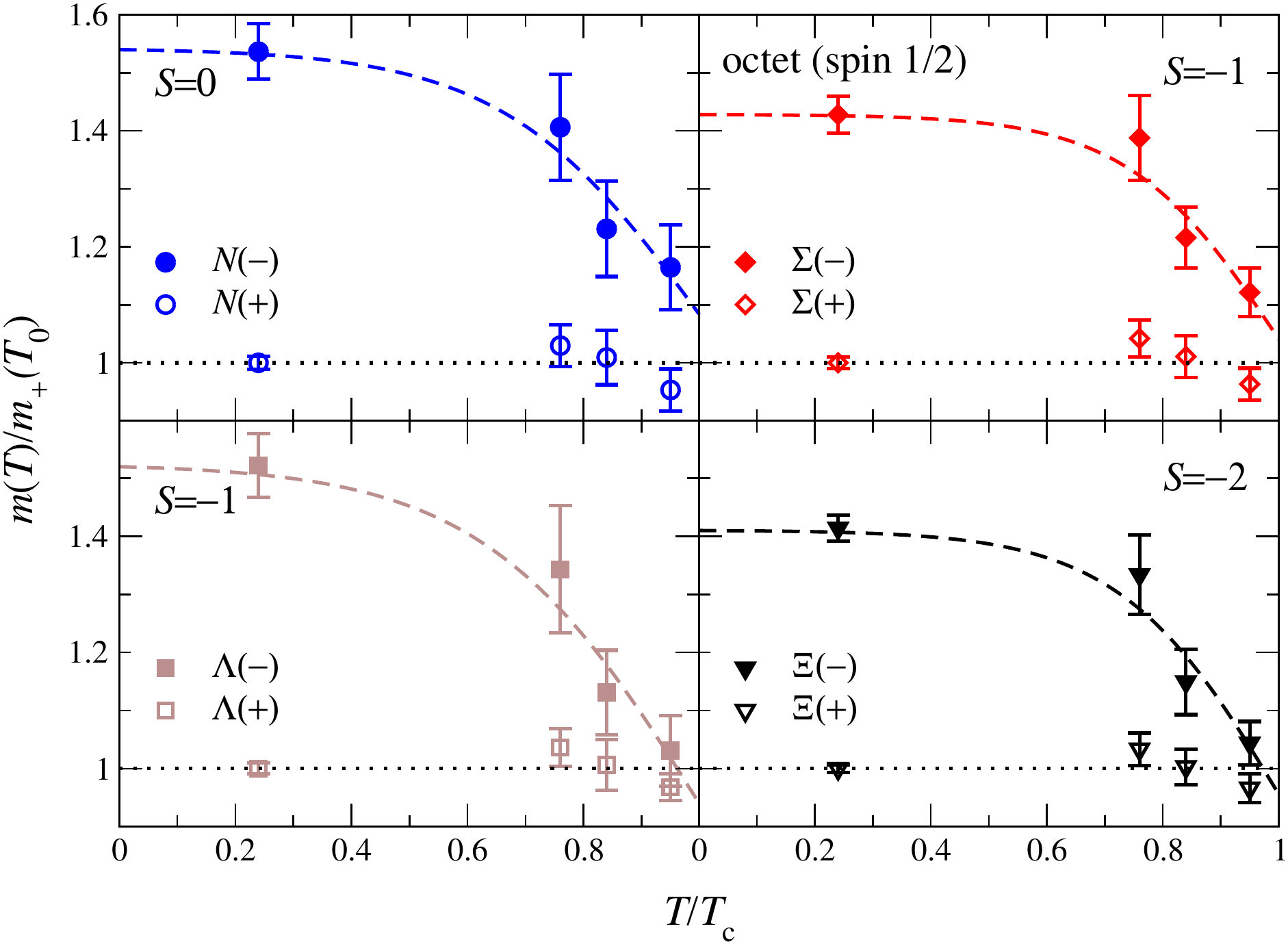}
  \includegraphics[width=6.6cm,clip]{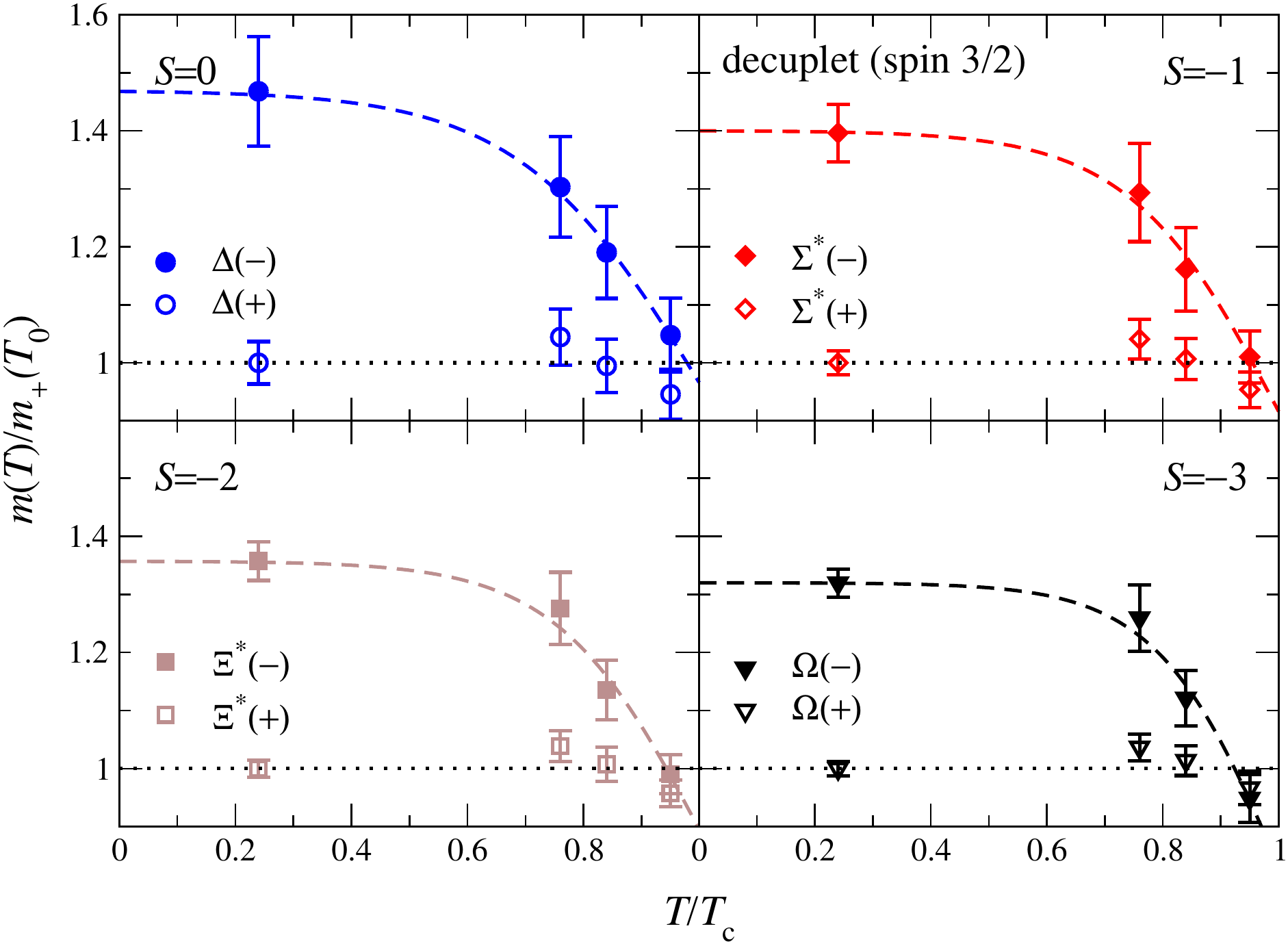}
  \caption{Groundstate masses obtained using exponential fits of the baryonic octet (left) and decuplet (right) correlators as in eq.(\ref{eq:Gsimple}) for temperatures below $T_c\,$. Positive-(negative-)parity
masses are represented by open (closed) symbols. The masses are normalised with the positive-parity groundstate mass at the lowest temperature, i.e.\ $m_{\pm}(T)/m_+(T_0)\,$, with $T_0=0.24T_c$.}
  \label{fig-2}
\end{figure}

The emergence of parity doubling can be seen also more directly in Fig.~\ref{fig-2}, in which the positive- and negative-parity groundstate masses are extracted below $T_c$ from the exponential fit of the correlator as in eq.(\ref{eq:Gsimple}). 
The groundstate masses in the positive-parity channel are approximately independent of the temperature,
whereas the groundstates in the negative-parity sector become lighter as the temperature is increased. This is found to be the case for all the baryons in the spin $1/2$ octet and spin $3/2$ decuplet.
This means that in-medium effects are stronger in the negative-parity channel and coincide with
the restoration of chiral symmetry. At the crossover temperature the groundstate parity partners become near-degenerate in mass, explaining why the $R$ factor drops around $T_c\,$. 

\begin{figure}[thb]
  \centering
  \includegraphics[width=6.4cm,clip]{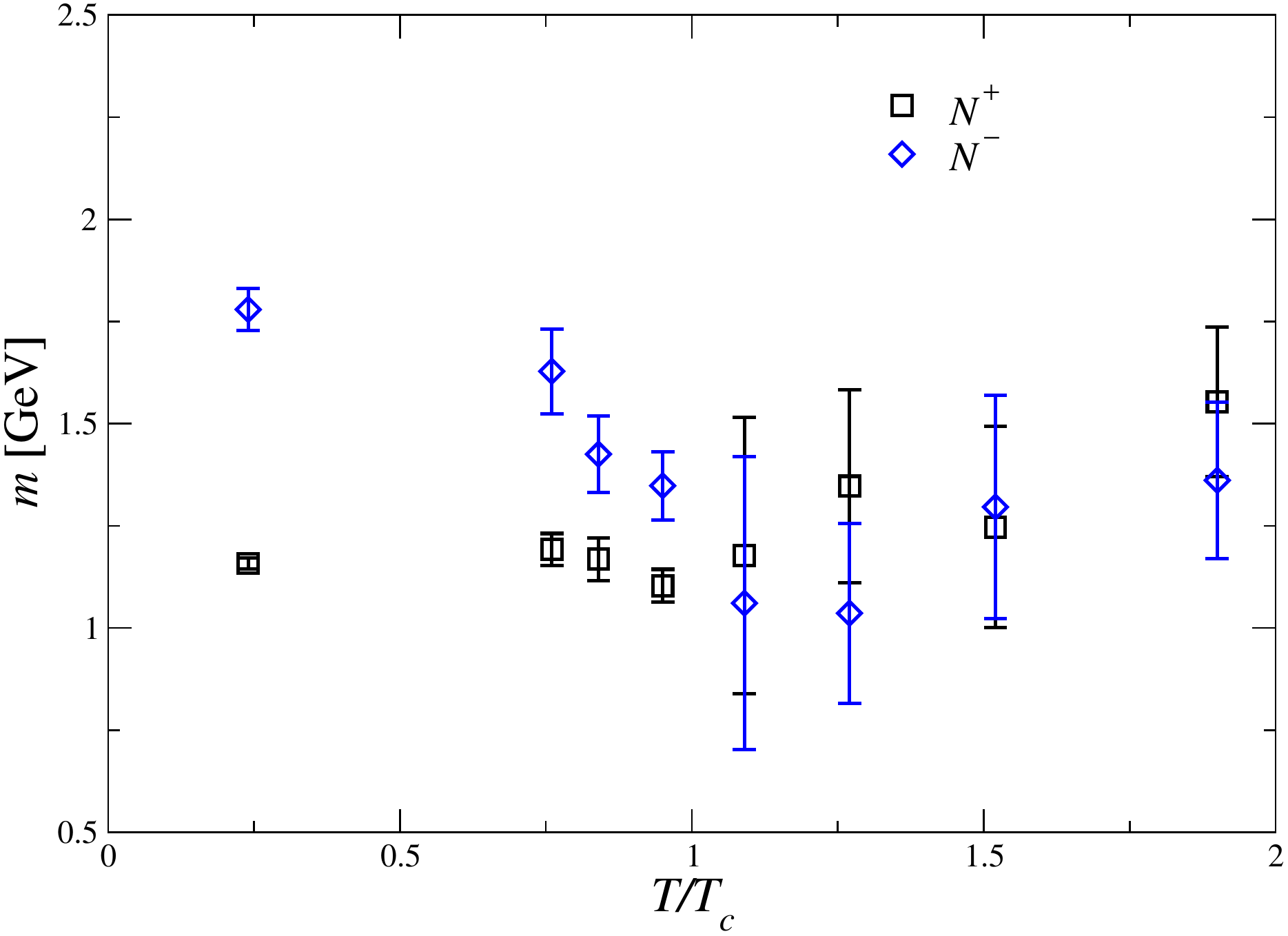}
  \includegraphics[width=6.4cm,clip]{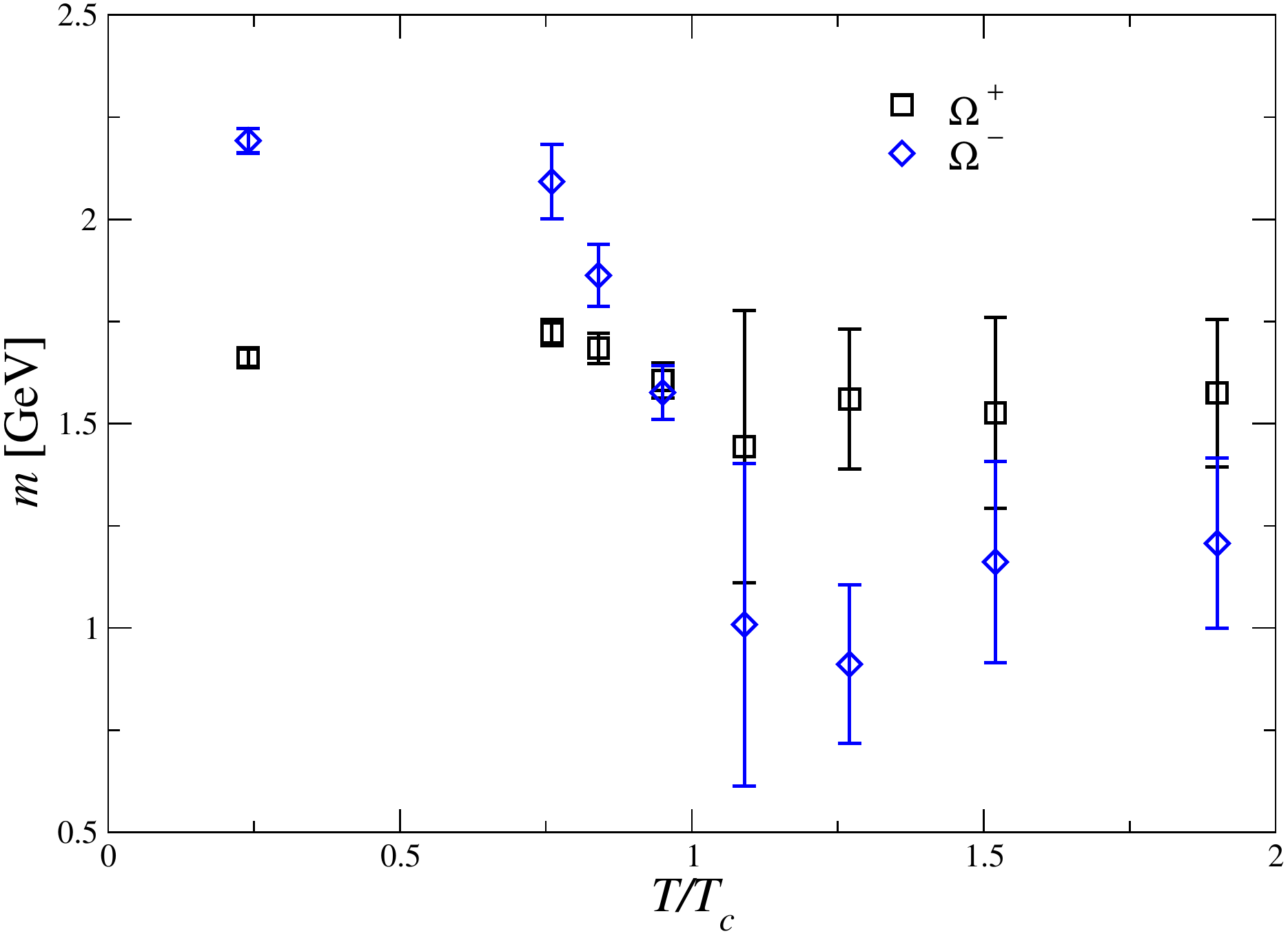}
  \caption{Nucleon (left) and $\Omega$-baryon (right) groundstate masses obtained using exponential fits of the correlator (see eq.\ref{eq:Gsimple}) both in the hadronic and quark-gluon plasma phase.}
  \label{fig-3}
\end{figure}

We also tried to extend the exponential fit above $T_c\,$, to check whether the groundstate signals remain in the high-temperature phase. Fig.~\ref{fig-3} shows that applying the same fitting procedure leads to a drastic increase in the error of the fitting parameters. The most likely interpretation is that the positive-and negative-parity groundstates have dissolved for temperatures higher than $T_c\,$. In Fig.~\ref{fig-3} only the masses of the nucleon and $\Omega$ baryon are shown, but the qualitative behaviour of the masses of the other octet and decuplet baryons is analogous.

The dashed lines in the plots of Fig.\ref{fig-2} correspond to a fit of the temperature dependence of the negative-parity masses between $T=0$ and $T=T_c$. We considered a fit function
\be\label{fitm}
m_-(T)=w(T,\gamma)\,m_-(0)+(1-w(T,\gamma))\,m_-(T_c)\:,
\ee
with the transition function $w(T,\gamma)=\tanh\left[(1-T/T_c)/\gamma\right]/\tanh[1/\gamma]\,$, such that $w(0,\gamma)=1$ and
$w(T_c,\gamma)=0\,$. The smaller the value of $\gamma$ the narrower the transition region, which is expected to
depend on the masses of the light quarks. The fits have been applied to all $8$ channels and we found that
$0.22\lesssim \gamma \lesssim 0.35$ and $0.85 \lesssim m_-(T_c)/m_+(0)\lesssim 1.1\,$. The major source of uncertainty resides in $m_-(T_c)\,$, since it
implicitly assumes that the groundstate survives at or close to the crossover temperature.

\section{Application to the Hadron Resonance Gas model}\label{sec:HRG}

\begin{figure}[thb]
  \centering
  \includegraphics[width=6.55cm,clip]{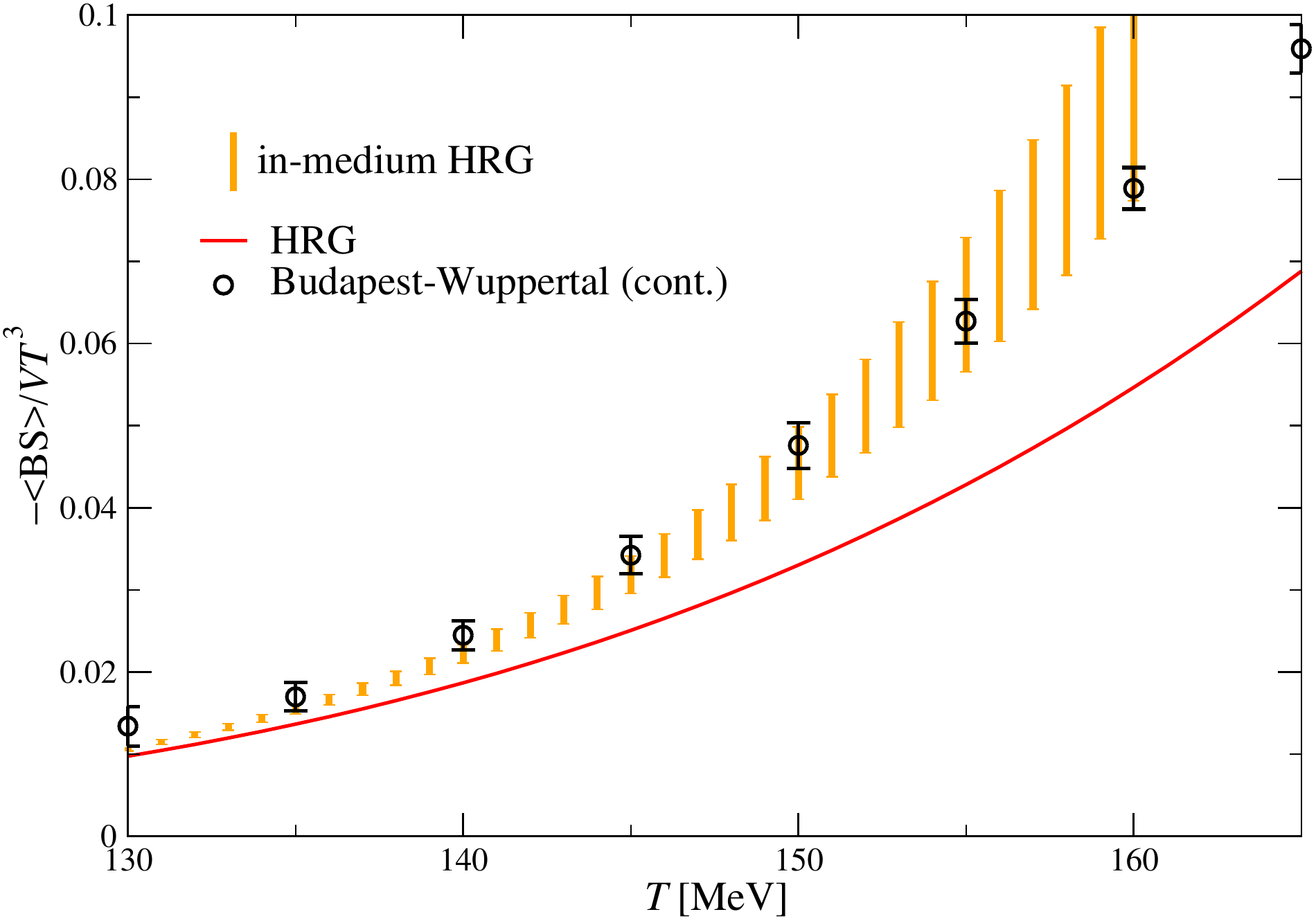}
  \includegraphics[width=7cm,clip]{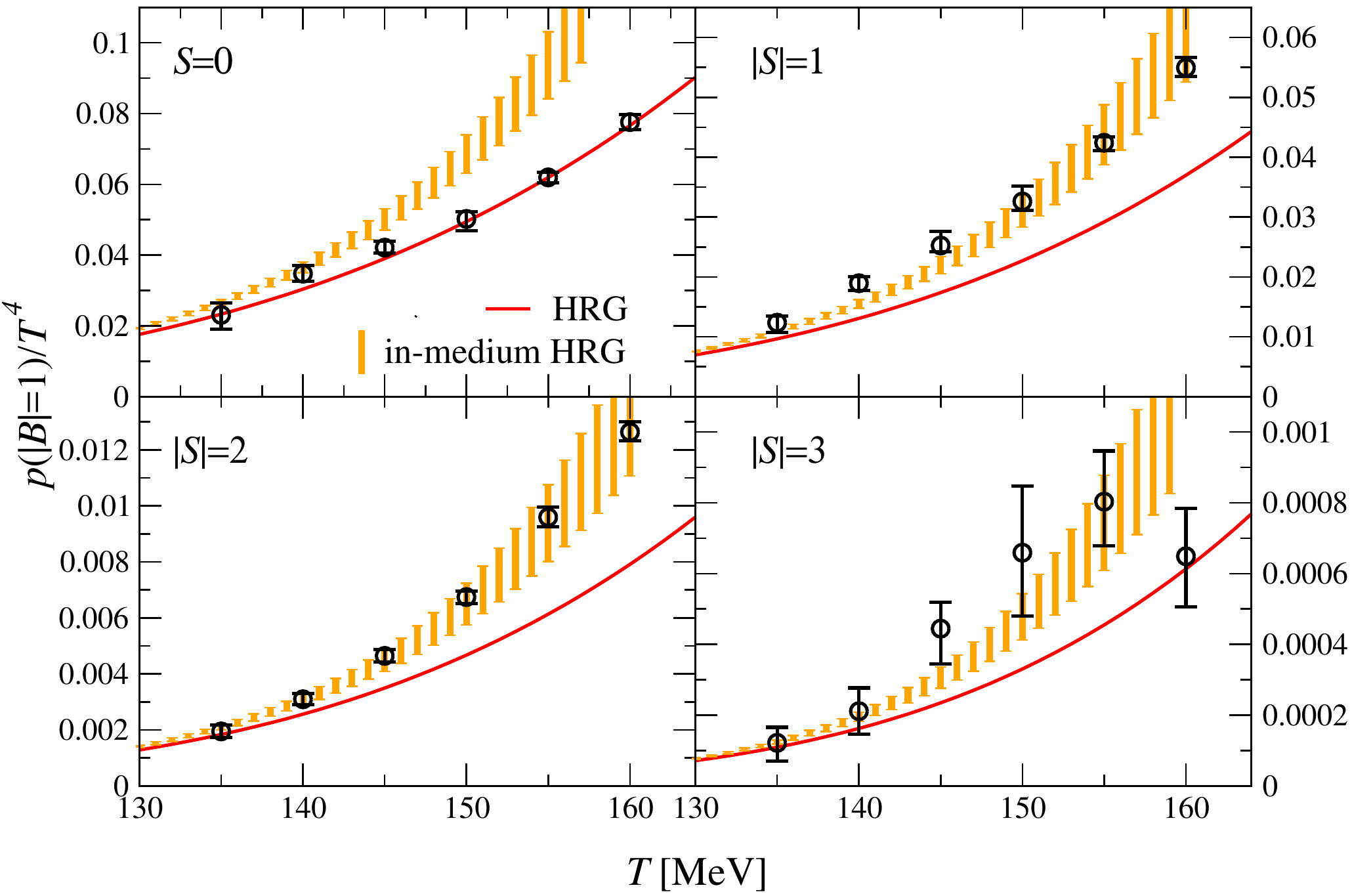}
  \caption{Left: Mixed fluctuation of baryonic (B) and strange (S) charge as defined in eq.(\ref{fluct}). Right: Contributions to the normalised pressure $P/T^4$ from the sectors with baryon number $|B|=1$ and strangeness $|S|=0,1,2,3\,$. Lattice data are taken from \cite{Borsanyi:2011sw,Bellwied:2015lba} and private communication (left) and from \cite{Alba:2017mqu} (right). The curves are obtained from the standard hadron resonance gas (HRG) and the in-medium HRG, using temperature- dependent negative-parity groundstate masses as in eq.(\ref{fitm}), with $\gamma=0.3\,$, $1< m_-(T_c)/m_+(0)<1.1$ and $T_c=155$ MeV.}
\label{fig-4}
\end{figure}

Here we show an application of the medium-dependent masses that might be of interest to heavy-ion phenomenology.
To do so we apply our in-medium modifications of the negative-parity groundstate masses to the hadron resonance gas (HRG) model.
The standard HRG uses vacuum masses, see for instance \cite{Alba:2017mqu} and references therein.
Our strategy is to keep the vacuum values for the positive-parity masses, which is supported by our results, and
modify the negative-parity groundstate masses according to eq.(\ref{fitm}). In our calculation we used the PDG2016 baryon masses
classified with $3$ and $4$ stars, up to $2.5$ GeV. For the in-medium modification we considered $\gamma=0.3$
and $1<m_-(T_c)/m_+(0)<1.1$ (the effect of varying $\gamma$ is negligible) and $T_c = 155$ MeV \cite{Borsanyi:2010bp}.
Fig.\ref{fig-4} shows both a fluctuation of conserved charges (in this case baryon number $B$ and strangeness $S$), defined as
\be\label{fluct}
\chi_{_{BS}}=\frac{1}{VT^3}\left.\frac{\partial^2\ln Z}{\partial\mu_B\partial\mu_S}\right|_{\mu=0}=\frac{1}{T^4}\sum_{i} B_iS_iP_i(\mu=0) \equiv \frac{1}{VT^3}\avg{BS}
\ee
and the single baryonic contributions to the pressure, according to the value of the strange number. In eq.(\ref{fluct}), $Z$ is the partition function, $P_i$ is the partial pressure and $\mu_{_{\rm X}}$ is the chemical potential associated with the conserved charge X. Note that $B=1$ for baryons.
The standard and in-medium HRG are contrasted with the lattice data of \cite{Alba:2017mqu}.
From the right panel of Fig.\ref{fig-4} one clearly sees that the standard HRG fits the lattice data only for $S=0\,$, whereas the in-medium HRG overshoots the data.
On the other hand, for $|S|=1,2,3$ the in-medium HRG describes the data much better than the standard one.
It would be interesting to see how this in-medium effect due to chiral symmetry restoration would compare with 
other modifications of the HRG that are currently under consideration \cite{Alba:2017mqu,Vovchenko:2016rkn},
for example, including the thermal width of the baryonic states when calculating thermodynamic quantities within the HRG model.

\clearpage
\section{Conclusions}\label{sec:conclusions}

In this work we considered octet and decuplet hyperons at non-zero temperature and found a clear emergence of parity doubling around the crossover temperature $T_c$ due to the restoration of chiral symmetry. This has been studied both at the level of the correlators ($R$ factor) and the groundstate masses. We showed that the positive- and negative-parity groundstate channels become near-degenerate around $T_c\,$, even in presence of the explicit breaking of chiral symmetry due to the finite mass of the $u$ and $d$ quarks. The effect of the heavier $s$ quark is visible in the emergence of parity doubling. The mass degeneracy originates via an in-medium modification of the negative-parity groundstate masses of all octet and decuplet baryons in the hadronic phase. That is, the groundstate mass of the negative-parity channels diminishes as the temperature increases, whereas the groundstate mass of the positive-parity channels remains unaffected (within the error bars) by temperature variations below $T_c$.\\
This in-medium effect might be relevant to heavy-ion phenomenology, in particular when applied to the hadron resonance gas model. We saw for example that the temperature dependence of the negative-parity groundstate masses reproduces quite well the lattice data regarding fluctuations of conserved charges ($\chi_{_{BS}}$) and partial pressures of hyperons (i.e.\ baryons with $|S|>0$). 
However, some additional effects seem to be relevant in the $S=0$ sector, where the standard HRG happens to reproduce the data of the thermodynamic quantities below $T_c$ already well. These additional effects might be related to the fact that one should take into account also the thermal width of the states and not only their mass.\\
Finally, parity doubling is considered in effective parity doublet models, in which it introduces a chirally invariant contribution to the baryonic masses \cite{PhysRevD.39.2805}. Our results can be used to further analyze/constrain these, see e.g.\ \cite{Benic:2015pia,Sasaki:2017glk,Mukherjee:2017jzi}.


\section*{Acknowledgments}

We thank Paolo Alba for discussions and Szabolcs Bors\'{a}nyi and Claudia Ratti for providing the lattice data shown in Fig.\ref{fig-4}.
This work has been supported by STFC grant ST/L000369/1, ICHEC, the Royal Society, the Wolfson Foundation and the Leverhulme Trust, and has been performed in the framework of COST Action CA15213 THOR.
We are grateful for the computing resources
made available by HPC Wales. This project used the DiRAC Blue Gene Q Shared
Petaflop system at the University of Edinburgh, operated by the Edinburgh
Parallel Computing Centre on behalf of the STFC DiRAC HPC Facility
(www.dirac.ac.uk). This equipment was funded by BIS National E-infrastructure
capital grant ST/K000411/1, STFC capital grant ST/H008845/1, and STFC DiRAC
Operations grants ST/K005804/1 and ST/K005790/1. DiRAC is part of the National
E-Infrastructure.

\clearpage
\bibliography{lattice2017}

\end{document}